\documentclass[prl,twocolumn,showpacs]{revtex4}
\usepackage{graphicx}% Include figure files
\usepackage{dcolumn}% Align table columns on decimal point
\usepackage{bm}% bold math

\begin{document}

\title{Brane induced gravity from asymmetric warped compactification}
% Force line breaks with \\

\author{Kazuya Koyama$^{1,2}$, Kayoko Koyama$^{3}$}
 
\affiliation{$^1$
Department of Physics, University of Tokyo 
7-3-1 Hongo, Bunkyo, Tokyo 113-0033, Japan\\
$^2$ Institute of Cosmology and Gravitation, Portsmouth 
University, Portsmouth, PO1 2EG, UK\\
$^3$ Department of Phyics and Astronomy, University of Sussex, 
Brighton, BN1 9QH, UK}

\date{\today}% It is always \today, today,

\begin{abstract}
We show that brane induced gravity can be realized as a low energy 
effective theory of brane worlds with asymmetric warped compactification.
A self-accelerating universe without cosmological constant on the brane 
can be realized in a model where one side of the bulk has finite volume, 
but the other side has infinite volume. 
The spin-2 perturbations for brane induced gravity and
asymmetric warped compactification models have the same spectrum at low
energies. For a de Sitter brane, the spin-2 graviton
has mass in the range $0<m^2 \leq 2H^2$, with $m^2=2H^2$
in the self-accelerating universe. 
\end{abstract}

\maketitle
%%%%%%%%%%%%%%%%%%%%%%%%%%%%%%%%%%%%%%%%%%%%%%%%
\noindent
{\it 1. Introduction}
\vspace{0.2cm}

%%%%%%%%%%%%%%%%%%%%%%%%%%%%%%%%%%%%%%%%%%%%%%%%
The discovery of the acceleration of universe has raised 
significant challenges for cosmology \cite{SN}. 
The simplest solution is 
the introduction of a tiny cosmological constant by hand. 
However, it requires an unacceptable fine-tuning. 
Recently, much efforts have been devoted to 
the attempt to modify Einstein theory of gravity. 
However, it is extremely difficult to modify gravity 
at large distances and, at the same time, provide the mechanism 
for cosmic acceleration \cite{T}. 

So far, one successful realization is known in the context 
of the brane-world scenario where standard model particles are
confined to the brane while gravity can propagate 
in higher-dimensional spacetime. Dvali, Gabadadze and Porrati
(DGP) \cite{DGP} proposed the brane induced gravity model 
where the 4D Einstein-Hilbert term is assumed 
to be induced on the brane. In this model, 4D gravity is 
recovered on short scales but gravity becomes 5D 
on large scales. Deffayet showed that the DGP model 
realized the accelerated expansion of the universe 
at late times without introducing the cosmological constant 
\cite{D}. 
The motivation for the induced Einstein-Hilbert term in the
action is that it arises from quantum effects of matter fields 
confined to the brane \cite{DGP}. 

We present a new way to realize the brane induced gravity model 
as an asymmetric warped compactification of the form discussed 
by Padilla \cite{Padilla1}.
Warped compactification is a mechanism to recover 4D gravity
on the brane via an extra-dimension that shrinks exponentially
away from the brane and so confines gravity \cite{RS}. 
Recent developments in string theory suggest that there may be many 
regions in the extra-dimension with different warped geometries, 
and with the observable universe on one of the D-branes  
between different warped geometries \cite{Warped}.

In this letter, we present a simple argument to show that
{\it the brane induced gravity model can be realized 
as a low energy effective theory of the asymmetric warped 
compactification model where one side of the brane confines 
the gravity but the other side does not}. 
\vspace{0.4cm}

%%%%%%%%%%%%%%%%%%%%%%%%%%%%%%%%%%%%%%%%%%%%%%%%%
\noindent
{\it 2. Brane Induced Gravity}
\vspace{0.2cm}

%%%%%%%%%%%%%%%%%%%%%%%%%%%%%%%%%%%%%%%%%%%%%%%%%
Let us consider the 5D action given by
\begin{eqnarray}
S &=& \int d^5 x \sqrt{-{}^{(5)\!}g} \left[ {1 \over 2 
\kappa^2} \left( {}^{(5)\!}R  +{12\over \ell^2} \right) \right] \nonumber\\
&& + \int d^4 x \sqrt{- {}^{(4)\!} g} 
\left( -\sigma + \frac{1}{2 \kappa_4^2} {}^{(4)\!} R \right).
\end{eqnarray} 
We assume that the tension is tuned $\sigma=6/\kappa^2 \ell$ so that 
the cosmological constant vanishes on the brane. The reflection symmetry across 
the brane is imposed. The original DGP model assumed $\sigma=0$ and 
$\ell \to \infty$. 

In this model, there is a characteristic scale called 
a cross over scale defined by
\begin{equation}
r_c= \frac{\kappa^2}{2 \kappa_4^2}.
\end{equation}
$r_c$ controls the strength of the brane induced gravity. 
In the following we assume $r_c \ll \ell$, which is the
regime relevant for self-accelerating models. 
The behavior of gravity on the brane is summarized in Fig.1 
\cite{Tanaka}.

\begin{figure}[t]
\centerline{
\includegraphics[width=9cm]{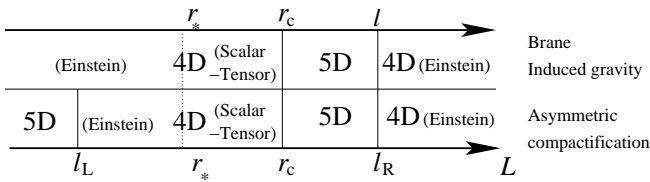}}
\caption{
Behavior of gravity outside a matter distribution with  
total mass $m$.
If we consider the scale $L < \ell$, gravity is not localized.
In the intermediate region $r_c < L < \ell$, gravity looks 5D.
4D Newton gravity is reproduced for $L < r_c$ due to the brane 
induced gravity, but linearized gravity is not described by 
Einstein theory because 4D gravity is recovered 
by massive modes which contain an extra scalar polarization.
However, there is a length scale given by 
$r_* = ( m \kappa_4^2 r_c^2/4 \pi)^{\frac{1}{3}}$ \cite{strong}. 
Below $r_*$ the non-linear interaction of the scalar mode 
effectively shields the extra scalar polarization. 
Then we recover 4D Einstein gravity for 
$L < r_*$. In the asymmetric warped compactirication model, 
there is a 5D gravity regime for $L<\ell_L$.}
\label{fig:scale}
\end{figure}

An interesting property of this model is that there is a solution 
for an accelerating universe without cosmological constant 
\cite{D,sahni}. 
The Friedmann equation on the brane is given by
\begin{equation}
\pm H = r_c H^2- \frac{\kappa^2}{6} \rho.
\end{equation}
The sign is related to the geometry of the bulk spacetime. 
We notice that in the $+$ branch, the universe approaches 
de Sitter spacetime at late times, i.e. $\rho \to 0$,
\begin{equation}
H = \frac{1}{r_c}.
\label{Hrc}
\end{equation}

In order to derive the above results, the junction condition
plays the crucial role. The junction condition at the brane
is given by
\begin{equation}
K^{\mu}_{\nu} = {\kappa^2 \over 2} 
\left( -{\sigma \over 3} \delta^{\mu}_{\nu} - 
\tilde{T}^{\mu}_{\nu} + {1 \over \kappa_4^{2}} 
{}^{(4)\!}\tilde{G}^{\mu}_{\nu} \right),
\end{equation}
where $K^{\mu}_{\nu}$ is the extrinsic curvature of the brane,
$^{(4)\!} G^{\mu}_{\nu}$ is Einstein tensor on the brane and 
$\tilde{T}^{\mu}_{\nu} = T^{\mu}_{\nu}-(1/3)T \delta^{\mu}_{\nu}$,
${}^{(4)\!}\tilde{G}^{\mu}_{\nu} ={}^{(4)\!}
G^{\mu}_{\nu}-(1/3)G \delta^{\mu}_{\nu}$. The 4D Einstein tensor
comes from induced Einstein-Hilbert term.
\vspace{0.4cm}

%%%%%%%%%%%%%%%%%%%%%%%%%%%%%%%%%%%%%%%%%%%%%%%%%%%%%%%%%%%
\noindent
{\it 3. Asymmetric Warped Compactification}
\vspace{0.2cm}

%%%%%%%%%%%%%%%%%%%%%%%%%%%%%%%%%%%%%%%%%%%%%%%%%%%%%%%%%%%
Next, let us consider the asymmetric warped compactification model 
described by the action \cite{Padilla1}
\begin{eqnarray}
S &=& \int_{M_R} d^5 x \sqrt{-{}^{(5)\!}g} \left[ {1 \over 2 
\kappa_R^2} \left( {}^{(5)\!}R  +{12\over \ell_R^2} \right) \right] \nonumber\\
&& +\int_{M_L} d^5 x \sqrt{-{}^{(5)\!}g} \left[ {1 \over 2 
\kappa_L^2} \left( {}^{(5)\!}R  +{12\over \ell_L^2} \right) \right]\nonumber\\
&& - \int d^4 x \sqrt{- {}^{(4)\!} g} \:\: \sigma, 
\label{action} 
\end{eqnarray} 
where $M_R, M_L$ are two AdS spacetimes with different
AdS curvature lengths $\ell_R, \ell_L$ (See Fig.2). The indices $R$ and $L$ 
are used to denote the variables in the right and left side 
of AdS spacetime respectively. 
We only consider the brane tension $\sigma$ and do not 
introduce an induced gravity term ${}^{(4)\!}R$ 
in the brane action. Instead of it, we have 
an asymmetry between the right and left side of the brane. 

We can explicitly show the equivalence of this model 
at low energies with the brane induced gravity model as follows. 
We first take $\ell_R \gg \ell_L$ and consider the physics 
on the scale $L$ on the brane with $L \gg \ell_L$. 
We solve the 5D Einstein equation for the bulk gravitational 
field in left side of the brane perturbatively in terms 
of the small parameter $(\ell_L/L)^2$. 
The solution for $K^{\mu}_{\nu,L}$ on the brane 
up to the first order is given by \cite{low}
\begin{equation}
K^{\mu}_{\nu,L}=\frac{1}{\ell_L} \delta^{\mu}_{\nu}
+\frac{\ell_L}{2} {}^{(4)\!}\tilde{G}^{\mu}_{\nu} + \chi^{\mu}_{\nu},
\end{equation}
where $\chi^{\mu}_{\nu}$ is an integration constant which should 
be determined by the conditions at Cauchy horion of the AdS spacetime. 
In the following we take $\chi^{\mu}_{\nu}=0$.
This assumption is appropriate for the
background dynamics, and we will consider
perturbations about the background in the following section. 
The appearance of Einstein
tensor manifests the localization of gravity in this model. 
This Einstein tensor is responsible for the recovery of 4D Einstein 
gravity in the model proposed by Randall and Sundrum \cite{RS} 
where the reflection symmetry across the brane is assumed. 

The junction condition for the asymmetric warped compactification
is given by
\begin{equation}
\kappa_R^{-2} K^{\mu}_{\nu,R} - \kappa_L^{-2} K^{\mu}_{\nu,L}
=-{\sigma \over 3} \delta^{\mu}_{\nu}-\tilde{T}^{\mu}_{\nu}.
\end{equation}
Using the solution for $K^{\mu}_{\nu,L}$ we derive 
the junction condition for $K^{\mu}_{\nu,R}$ as 
\begin{equation}
K^{\mu}_{\nu, R}=\kappa_R^2 \left(-\frac{\sigma_{\rm eff}}{3} -\tilde{T}^{\mu}_{\nu}
+\frac{\ell_L}{2 \kappa_L^2} {}^{(4)\!}\tilde{G}^{\mu}_{\nu}  \right).
\end{equation}
where $\sigma_{{\rm eff}}=\sigma-3/\kappa_L^2 \ell_L$. 
This is completely the same as the junction condition in the brane
induced gravity model
except for factor $2$ that comes from the reflection symmetry in the brane
induced gravity model.
We find that 
{\it the localization of the gravity in the left side plays the same role
as brane induced gravity}.
We can identify the coupling constant $\kappa_4^2$ and
the cross-over scale $r_c$ as 
\begin{equation}
\displaystyle{
\kappa_4^2 =2 \kappa_L^2 \ell_L^{-1}~ ,~~~~
r_c= \frac{\ell_L  \kappa_R^2}{2 \kappa_L^2}.}
\label{rc}
\end{equation}
In order to ensure $\ell_L \ll r_c \ll \ell_R$, we must choose 
the parameters so that $\ell_L \kappa_R^2 \ll \ell_R \kappa_L^2$ 
and $\kappa_L^2  \ll \kappa_R^2$.

Padilla studied the behavior of gravity in asymmetric brane \cite{Padilla1} 
and noticed that the behavior of gravity 
is pretty much the same as the brane induced gravity model 
studied in Ref. \cite{Tanaka} (See Fig.1). Indeed,  
we can explicitly check that the results of \cite{Padilla1} 
can be derived from the results of \cite{Tanaka} in the 
brane induced gravity model using the identification 
(\ref{rc}).  

The cosmological solutions can be easily obtained. We tune the 
tension $\sigma = 3/\kappa_L^2 \ell_L + 3/\kappa_R^2 \ell_R$ so that 
the cosmological constant on the brane vanishes. 
Then we have the same Friedmann equation (3). 
Actually we have the late time accelerating universe \cite{Padilla1}
\begin{equation}
H=\frac{1}{r_c} = \frac{2 \kappa_L^2}{\ell_L \kappa_R^2}.
\end{equation}
\vspace{0.1cm}

%%%%%%%%%%%%%%%%%%%%%%%%%%%%%%%%%%%%%%%%%%%%%%%%%%%%%%%
\noindent
{\it 4. Linear Perturbations and Mode Analysis}
\vspace{0.2cm}

%%%%%%%%%%%%%%%%%%%%%%%%%%%%%%%%%%%%%%%%%%%%%%%%%%%%%%%
We now show that the relation between brane induced gravity 
and asymmetric warped compactification in the background 
dynamics extends also to the linear perturbations
about Minkowski brane and de Sitter brane. We study the structure
of the mass spectrum of the discrete and continuous modes in both 
the brane induced gravity model and the asymmetric warped 
compactification model. One can see the brane induced gravity 
is actually recovered as a low energy limit of the 
asymmetric warped compactification.

We write the background metric in the bulk as
\begin{equation}
ds^2=dy^2 + a(y)^2 \gamma_{\mu \nu} dx^{\mu} dx^{\nu},\quad 
\end{equation}
where $\gamma_{\mu \nu}$ is the metric of Minkowski spacetime
or de Sitter spacetime. We assume the brane is located at 
$y=0$.
Let us consider the perturbations $\gamma_{\mu \nu}+ h_{\mu \nu}$. 
Imposing the transverse-traceless condition $\nabla^{\mu} h_{\mu \nu}=
h^{\mu}_{\mu}=0$ and 
%, the wave equation in the bulk is given by
%\begin{equation}
%h''_{\mu \nu} + 4 {a' \over a} h'_{\mu \nu} +{1 \over a^2} 
%(\nabla^2 -2H^2) h_{\mu \nu} =0,
%\label{h}
%\end{equation}
%where $H$ is the Hubble scale on the brane and $\nabla$ is 
%the covariant derivative with respect to $\gamma_{\mu \nu}$.
using the separation of variables, i.e. the mode expansion 
$h_{\mu \nu} = \int dm~ e_{\mu \nu}(x) F_m(y)$, 
one can rewirte the wave equation in the bulk 
as the Shr\"{o}dinger equaiton 
\begin{equation}
-\frac{1}{2} \frac{d^2}{d z^2}\Psi_m +V_{\rm eff} \Psi_m 
= \frac{1}{2}m^2 \Psi_m,
\label{psi}
\end{equation}
where $\Psi_m =a^{3/2}F_m$ and $z$ is a conformal coordinate $a(y)dz =dy$.
In order to have the normalizable mode we must impose the condition
$\int_0^{\infty} dy a^2 F_m^2 < \infty$ for discrete modes and 
$\int_0^{\infty} dy a^2 F_m F_{m'} =\delta(m-m')$ for continuous 
massive modes.
\vspace{0.3cm}

\noindent
{\it 4.1 Mincowski Brane}
\vspace{0.2cm}

\begin{figure}[t]
\centerline{
\includegraphics[width=9cm]{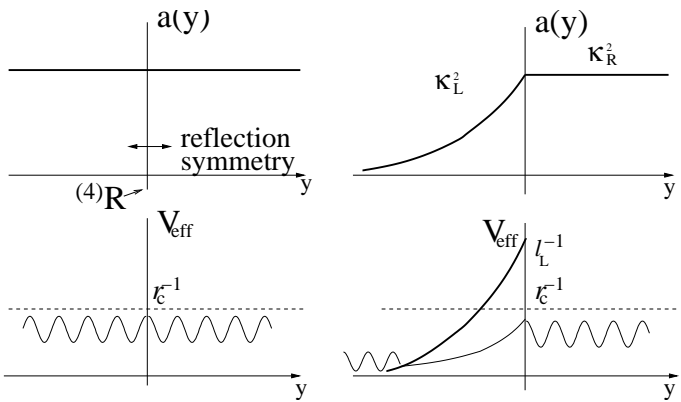}}
\caption{Warp factor and effective potential for Minkowski brane.
In the brane induced gravity model, $\sigma=0$, the warp factor 
and the effective potential are given by $a=1$ and $V_{{\rm eff}}=0$, 
respectively, for $-\infty < y < \infty$ if we take $\ell \to \infty$.
In the asymmetric warped compactification model, $\sigma >0$ and 
$a_L = e^{y/\ell_L},~
{V_{{\rm eff}}}_{\rm L} \propto {\rm e}^{2y/\ell_L} /\ell_L^2 ~
{\rm for} ~y<0 ~ , ~
 a_R = 1,~{V_{{\rm eff}}}_{\rm R}=0~{\rm for} ~y>0$. 
}
\label{fig:minkowski}
\end{figure}

First we investigate the Mikonwski brane case, $H=0$ (See Fig.2).
In the brane induced gravity model, 
the junction condition on the brane is given by
\begin{equation}
F_m'(0)=-m^2 r_c F_m(0).
\end{equation}
Note that the 0-mode solution that satisfies this boundary condition 
is not normalizable and so it is excluded from the spectrum.
The solution for $F$ on the brane is given by \cite{msol}
\begin{equation}
\vert F_m(0) \vert^2 =\vert N \vert^2 \frac{4}{1+ m^2 r_c^2},
\end{equation}
where $N$ is the normalization factor. For $m > 1/r_c$, 
the contribution of massive modes are suppressed, so 
4D gravitational interactions are reproduced on the scale $L < r_c$. 

In the asymmetric warped compactification model (See Fig.2),
from Eq.(\ref{psi}) the mode functions $F_{L m}(y)$ 
in the left side AdS bulk are given by \cite{RS}
\begin{equation}
F_{L m}(y) 
= \frac{ e^{2 y/\ell_L} H_2^{(1)}(m \ell_L e^{2 y/\ell_L})}{H_2^{(1)}(m \ell_L)},
\end{equation}
where $H_2^{(1)}$ is a Hankel function of the first kind and we imposed 
a no-incoming radiation condition at the Cauchy horizon of AdS spacetime.
The derivative of $F_{L m}$ can be written as 
\begin{equation}
F_{L m}'(0)=-\left(\frac{m^2 \ell_L}{2} + \frac{m^2 \ell_L}{2}
 \frac{H_0^{(1)}(m \ell_L)}{H_2^{(1)}(m \ell_L)} \right).
\label{derivL}
\end{equation}
At low energies $m \ll 1/\ell_L $, we can neglect the second term in 
Eq.(\ref{derivL}). From the junction condition
$\kappa_R^{-2} F_{R m}' = \kappa_L^{-2} F_{L m}'$ , we find
\begin{equation}
F_{R m}'(0)= -\frac{\ell_L \kappa_R^2}{2 \kappa_L^2} m^2 F_{R m}(0) 
       = - m^2 r_c F_{R m}(0).
\end{equation}
Then, we have the same modes in the right side bulk 
as the brane induced gravity model. The induced gravity 
effect here is originated from the confinement of nearly massless 
modes around the brane in the left side AdS bulk. 

On the other hand, the second term in the right hand side 
of Eq.(\ref{derivL}) becomes important at high energies $m > 1/\ell_L$. 
We do not have an induced gravity effect and the 4D theory is modified. 
It is interesting to note that, in AdS bulk spacetime, 
the modes with a finite Kaluza-Klein mass are quasi-bound 
states because it can decay into infinity $y=-\infty$ 
due to tunnelling (See Fig.2). These phenomena give clear differences 
between the brane induced model and asymmetric warped 
compactification model at high energies. 
\vspace{0.3cm}

\noindent
{\it 4.2 de Sitter Brane}
\vspace{0.2cm}

Next we investigate the de Sitter solution that includes the 
self-accelerating solution (See Fig.3). 
We start with the brane induced gravity model, with
$V_{\rm eff}=9 H^2/4$ and $a_{+}=1 + H|y|$. In de Sitter spacetime, 
continuous massive modes start from $m^2 =9 H^2/4$. 
Again, the 0-mode solution with $m=0$
that satisfies the boundary condition is not normalizable. 
Continuous massive modes are too heavy to reproduce the 4D
gravity on super-horizon scales. However unlike the Minkowski 
brane case, Eq. (\ref{psi}) shows that there exists 
a normalizable discrete mode in de Sitter case 
that is given by
\begin{equation}
F_m(y)=a(y)^{-\frac{3}{2}-\sqrt{\frac{9}{4}-\frac{m^2}{H^2}}}.
\end{equation}
Imposing the junction condition, we find one discrete mode 
with mass 
\begin{equation}
\frac{m_d^2}{H^2} =\frac{1}{(H r_c)^2} \left(3 (H r_c)-1 \right).
\label{21}
\end{equation}
We should remember that the Hubble parameter is $H \geq 1/r_c$. Then 
we find $0< m_d^2 \leq 2 H^2$. 
For large $H r_c$ where induced gravity effect is large, 
we have a nearly massless mode $m_d^2 \to 0$. 
This mode is responsible for the recovery of 4D gravity 
for $1 \ll H r_c$ on super-horizon scales. The self-accelerating
universe is a de Sitter solution with $H r_c =1$, then 
Eq. (\ref{21}) shows that $m_d^2= 2 H^2$.

\begin{figure}
\centerline{
\includegraphics[width=9cm]{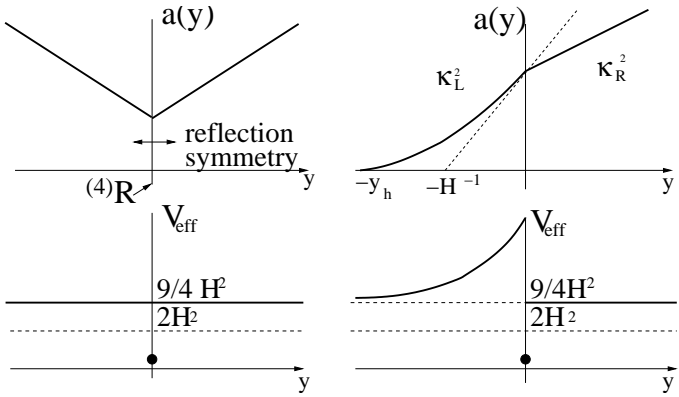}}
\caption{Warp factor and effective potential
for de Sitter brane.
The warp factor in the brane induced gravity models is given 
by $a_{+}(y)=1 + H \vert y \vert$ in $+$ branch.
For asymmetric compactification $
a_R(y) = 1+Hy ~{\rm for} ~ y>0 ~ , ~ 
a_L(y) = H \ell_L \sinh \left({y_h+y \over \ell_L}\right) 
~{\rm for}~ y<0,$
where $y_h$ is determined by $\sinh( y_h/\ell_L) = 1/H \ell_L$.
}
\label{fig:desitter}
\end{figure}

Let us now consider the asymmetric compactification model. 
We assume $\ell_R \to \infty$ for simplicity. 
The normalizable solutions for $0 < m^2 < 9/4 H^2$ 
in the left side of the bulk can be solved as
\begin{equation}
F_L(y) 
= \frac{
\sinh^{-2} \left({y_h + y \over \ell_L}\right)  
Q^{2}_{-\frac{1}{2}-\nu} \left( \coth \left({y_h + y \over \ell_L}\right) \right)}
{\sinh^{-2} \left( {y_h \over \ell_L} \right) 
Q^{2}_{-\frac{1}{2}-\nu} \left( \coth \left({y_h \over \ell_L}\right) \right)} ~, 
\end{equation}
where $\nu^2 =9/4-m^2/H^2$ and $Q^{\alpha}_{\beta}$ are associated 
Legendre functions of the second kind. 
At low energies $H \ell_L \ll 1$, we have 
$F_L' \sim -m^2 \ell_L/2$, thus we have the 
same boundary condition for $F_R$ and the same mass for the discrete mode 
in the right side of the brane as the brane induced gravity model. 

Note that the mass of the discrete 
mode is nearly 0 even at high energies $H > 1/\ell_L$. 
Thus on super-horizon scales, 4D physics is recovered 
even at high energies.  This is because
at high energies the warp factor approaches 
$a_L(y) \to a_-(y)=1+H y$ for $y<0$. 
The behavior of the left side bulk solution is the same as 
the $-$ branch solution in the DGP model while the right side one
behaves like the $+$ branch solution in the DGP model. 
Although the right side of the bulk volume is infinite, 
the left side of the bulk has finite volume, due to 
the horizon in the bulk $y=-y_h \to -H^{-1}$. 
Then a normalizable 0-mode appears in the  
left side. This supports a nearly massless mode and one
can recover 4D gravity on super-horizon scales. 

We should emphasize that the mass of the discrete mode lies
in the range $0< m_d^2 \leq 2 H^2$ for the spin-2 graviton.
Spin-2 massive graviton contains an extra scalar polarization. 
In 4D massive gravity theory where 
a Pauli-Fierz mass terms is introduced by hand \cite{PF}, 
this scalar polarization 
becomes a ghost if the mass lies in the range 
$0 < m^2 <2H^2$ \cite{Higuchi}. 
However, this conclusion depends on the introduction 
of an explicit mass term in the lagrangian and 
we cannot extrapolate this result to brane world models. 
In addition to spin-2 graviton, a spin-0 mode called the radion could be 
physical in the brane world and this raises another danger of a possible 
ghost \cite{strong}. 
A detailed analysis of gravity in de Sitter solution is needed to conclude 
that we have a consistent realization of massive gravity in the de Sitter 
brane. This is a crucial question to be addressed to realize the 
self-accelerating universe consistently \cite{KK}.  

We would like to thank R. Maartens for discussions and 
a careful reading of this manuscript. 
KK is supported by JSPS.


\begin{references}

\bibitem{SN}
S. Perlmutter et. al., 
Astrophys. J. {\bf 517} (1999) 565.

\bibitem{T}
S.~M.~Carroll, V.~Duvvuri, M.~Trodden and M.~S.~Turner,
Phys.\ Rev.\ {\bf D70} (2004) 043528;
N.~Arkani-Hamed, H.~C.~Cheng, M.~A.~Luty and S.~Mukohyama,
JHEP {\bf 0405} (2004) 074.

\bibitem{DGP}
G.~R.~Dvali, G. Gabadadze and M. Porrati, 
Phys. Lett. {\bf B485} (2000) 208.

\bibitem{D}
C. Deffayet, Phys. Lett. {\bf B502} (2001) 199.

\bibitem{Padilla1}
A. Padilla, Class. Quantum Grav. {\bf 22} (2005) 681; 
hep-th/0410033.

\bibitem{RS}
L. Randall and R. Sundrum, Phys. Rev. Lett. {\bf 83} (1999) 4690. 

\bibitem{Warped}
S.~B.~Giddings, S.~Kachru and J.~Polchinski,
Phys. Rev. {\bf D66} (2002) 106006.

\bibitem{Tanaka}
T. Tanaka, Phys. Rev. {\bf D69} (2004) 024001.

\bibitem{strong}
M.~A.~Luty, M. Porrati and R. Rattazzi, JHEP {\bf 0309} (2003) 029.

\bibitem{sahni}
V. Sahni and Y. Shtanov,  JCAP {\bf 0311} (2003) 014.

\bibitem{low}
S.~Kanno and J.~Soda,
Phys.\ Rev.\ {\bf D66} (2002) 043526.

\bibitem{msol}
G.~R.~Dvali, G.~Gabadadze, M.~Kolanovic and F.~Nitti,
Phys.\ Rev.\ {\bf D64} (2001) 084004.

\bibitem{PF}
M. Fierz and W. Pauli, Proc. Roy. Soc. {\bf 173} (1939) 211.

\bibitem{Higuchi}
A. Higuchi, Nucl. Phys. {\bf B282} (1987) 397.



\bibitem{KK}
K. Koyama, in preparation.


\end{references}
\end{document}